\documentclass{bmcart}
\usepackage{algorithm}
\usepackage{algorithmic}
\usepackage{booktabs} 
\usepackage{graphicx}
\usepackage{epsfig}
\usepackage{booktabs} 
\usepackage{threeparttable}
\usepackage{colortbl} 
\usepackage{color}
\usepackage{textcomp}
\usepackage{soul} 
\usepackage{amsmath}
\DeclareMathOperator*{\argmax}{argmax} 

\usepackage{comment}
\usepackage{multirow}
\usepackage{algorithm} 
\usepackage[utf8]{inputenc} 

\startlocaldefs
\endlocaldefs

\begin{document}

\begin{frontmatter}

\begin{fmbox}
\dochead{Methodology}

\title{edge2vec: Representation learning using edge semantics for biomedical knowledge discovery}


\author[addressref={aff1},  email={gao27@indiana.edu}]{\fnm{Zheng} \snm{Gao}}
\author[addressref={aff2},
email={ganfu@microsoft.com}]{\fnm{Gang} \snm{Fu}}
\author[addressref={aff3},
email={ouyangcp@gmail.com}]{\fnm{Chunping} \snm{Ouyang}}
\author[addressref={aff1},
email={stsutsui@indiana.edu}]{\fnm{Satoshi} \snm{Tsutsui}}
\author[addressref={aff1},
email={liu237@indiana.edu}]{\fnm{Xiaozhong} \snm{Liu}}
\author[addressref={aff1,aff2,aff4},
email={jejyang@iu.edu}]{\fnm{Jeremy} \snm{Yang}}
\author[addressref={aff1},
email={chris.r.gessner@gmail.com}]{\fnm{Christopher} \snm{Gessner}}
\author[addressref={aff5},
email={brian@d2discovery.com}]{\fnm{Brian} \snm{Foote}}
\author[addressref={aff1,aff5},
email={djwild@indiana.edu}]{\fnm{David} \snm{Wild}}
\author[addressref={aff1,aff5},
email={dingying@indiana.edu}]{\fnm{Ying} \snm{Ding}}
\author[addressref={aff6},corref={aff6},email={yuqi@sxmu.edu.cn}]{\fnm{Qi} \snm{Yu}}

\address[id=aff1]{
  \orgname{School of Informatics, Computing and Engineering, Indiana University},
  \city{Bloomington, IN},
  \cny{USA}            
}
\address[id=aff2]{
    \orgname{Microsoft Corporation},
    \city{Seattle, Washington},
    \cny{USA}
}

\address[id=aff3]{
    \orgname{University of South China},
    \city{Hengyang, Hunan},
    \cny{China}
}

\address[id=aff4]{
    \orgname{School of Medicine, University of New Mexico},
    \city{Albuquerque, NM},
    \cny{USA}    
}
\address[id=aff5]{
    \orgname{Data2Discovery, Inc.},
  \city{Bloomington, IN},
  \cny{USA}
}
\address[id=aff6]{
    \orgname{School of Management, Shanxi Medical University},
    \city{Taiyuan,  Shanxi},
    \cny{China}
}
\begin{artnotes}
\end{artnotes}

\end{fmbox}

\begin{abstractbox}

\begin{abstract}
\parttitle{Background}
Representation learning provides new and powerful graph analytical approaches and tools for the highly valued data science challenge of mining knowledge graphs. Since previous graph analytical methods have mostly focused on homogeneous graphs, an important current challenge is extending this methodology for richly heterogeneous graphs and knowledge domains. The biomedical sciences are such a domain, reflecting the complexity of biology, with entities such as genes, proteins, drugs, diseases, and phenotypes, and relationships such as gene co-expression, biochemical regulation, and biomolecular inhibition or activation.  Therefore, the semantics of edges and nodes are critical for representation learning and knowledge discovery in real world biomedical problems.
\parttitle{Results}
 In this paper, we propose the edge2vec model, which represents graphs considering edge semantics. An edge-type transition matrix is trained by an Expectation-Maximization approach, and a stochastic gradient descent model is employed to learn node embedding on a heterogeneous graph via the trained transition matrix. edge2vec is validated on three biomedical domain tasks: biomedical entity classification, compound-gene bioactivity prediction, and biomedical information retrieval. Results show that by considering edge-types into node embedding learning in heterogeneous graphs, edge2vec significantly outperforms state-of-the-art models on all three tasks. 
 \parttitle{Conclusions}
 We propose this method for its added value relative to existing graph analytical methodology, and in the real world context of biomedical knowledge discovery applicability.    
\end{abstract}

\begin{keyword}
\kwd{knowledge graph}
\kwd{heterogeneous network}
\kwd{biomedical knowledge discovery}
\kwd{representation learning}
\kwd{graph embedding}
\kwd{node embedding}
\kwd{edge semantics}
\kwd{applied machine learning}
\kwd{data science}
\kwd{linked data}
\kwd{semantic web}
\kwd{network science}
\kwd{systems biology}
\end{keyword}


\end{abstractbox}
%

\end{frontmatter}

\section*{Background}
\subsection*{Introduction}
The knowledge graph (KG) has become the preferred data model for complex knowledge domains. Accordingly Wilcke et al. published: "The knowledge graph as the default data model for learning on heterogeneous knowledge."\cite{wilcke2017kgdatamodel}. Biology and biomedical knowledge is complex and involves a plethora of entity and association types, hence is particularly suited to heterogeneous graph methodology. From such a KG, statistical knowledge can be inferred, for example, probabilistic associations between genes and phenotypic traits.  In KG terms, node and edge semantics are varied and critical for precise representation of the knowledge. Methods which consider surrounding node and edge contexts support a rich and combinatorially expanding feature set. KG embedding connotes representation of entities as computable feature vectors amenable to machine learning (ML) methods\cite{goodfellow2016deep,cai2018grembed}.  As both KG and ML methodology advances, the issues of embedding,  representation and vectorization become crucial, as signaled by related research activity spanning computing, natural and social sciences\cite{cai2018grembed}. Deep learning is a powerful approach for representation learning on large graphs and datasets. Multi-layer\ deep neural networks entail transformations from input raw data to layered representations obviating the need for feature engineering up front. Instead a set of continuous, latent features (representations) are learned which, in the graph use case, encode localized structural topology around a given node facilitating prediction tasks based on network structure. 

Previous work has focused on using neural network learning models to generate node embeddings for graphs such as DeepWalk\cite{perozzi2014deepwalk}, LINE\cite{tang2015line}, and node2vec\cite{grover2016node2vec}. However these models were designed for homogeneous networks, which means that they do not explicitly encode information related to the types of nodes and edges in a heterogeneous network. Recently, metapath2vec\cite{dong2017metapath2vec} was proposed by incorporating metapaths with node semantics for node embedding learning. However, this approach has several drawbacks: 1) domain knowledge is required to define metapaths and those mentioned in \cite{dong2017metapath2vec} are symmetric paths which are unrealistic in many applications; 2) metapath2vec does not consider edge types rather only node types; and 3) metapath2vec can only consider one metapath at one time to generate random walk, it cannot consider all the metapaths at the same time during random walk. On another related track, which might be termed biomedical data science (BMDS), previous work has employed KG embedding and ML methodology with the focus on applicability and applications such as compound target bioactivity\cite{chen2012assessing, seal2015randomwalk} and disease-associated gene prioritization\cite{himmelstein2015}. Yet other efforts have simply employed off-the-shelf ML toolkits (e.g. Scikit-learn, WEKA) and methods to address biomedical informatics prediction challenges. 

To address the above problems, edge2vec was developed to consider edge semantics when generating node sequence using a random walk strategy.  An edge-type transition matrix is defined to improve representation of node "context" and  designed with an Expectation-Maximization (EM) model. In the maximization step, we use the transition matrix to generate node sequences based on random walk in a heterogeneous graph. In the expectation step, we use the generated node ‘context‘ from node embeddings as feedback to optimize the transition matrix. We also use a skip-gram sampling strategy to select partial nodes for the EM approach to make the edge2vec model run on large-scale networks to learn node embeddings in a more efficient way. In the end, the topologically similar nodes (with similar sub-structures or located near each other in the network) are with similar emebeddings; the semantically similar nodes (with same node-types or logistically related attributes) are with similar embeddings.

Within biomedicine, the sciences involved in drug discovery are diverse. Drug efficacy and safety depend on calibrated modulation of complex, interrelated biomolecular pathways and targets. Prediction of compound-target bioactivity, normally non-covalent binding, remains high-challenge and high-value, both for generating novel drug leads and hypotheses, and for elucidating the mechanism of action for known compounds and drugs. With this rich knowledge domain as context, in this paper, we apply edge2vec on Chem2Bio2RDF\cite{chen2010chem2bio2rdf}, a highly heterogeneous graph integrating over 25 biomedical and drug discovery datasets.

The contribution of our work is threefold.

\begin{itemize}
\item We define an edge-type transition matrix to represent network heterogeneity. The calculation of the matrix is mainly based on the path similarity of different edge-types.

\item We develop an EM model to  train a transition matrix via random walks on a heterogeneous graph as a unified framework and employ a stochastic gradient descent (SGD) method to learn node embedding in an efficient manner. The learned node vector can include not only the topological information of network structure, but also the edge type information, which indicates different relationships among nodes.

\item We evaluate our model in the drug discovery domain by predicting drug-target associations using the highest available quality datasets as ground truth. Validation of the edge2vec model is addressed via three prediction tasks, all realistic biomedical discovery use cases. Validation results indicate that edge2vec adds value relative to existing methodology for drug discovery knowledge discovery.
\end{itemize}

In the following sections, first, we introduces edge2vec and its importance; second, we discusses related work about node embedding learning as well as heterogeneous network analysis; third, we explains edge2vec; fourth, we evaluates edge2vec based on later drug discovery; fifth, we illustrates two case studies to visualize edge2vec results, And in the end we concludes and points out future work.

\subsection*{Related Work}
\textbf{Network Representation:}
Network representation is useful in a variety of applications such as network classification\cite{bhagat2011node,sen2008collective}, content recommendation\cite{fouss2007random,yu2014personalized,gao2018end}, 
community detection\cite{gao2017personalized,liu2016comparing,zhang2016others} and link prediction\cite{liben2007link}. Networks are easily and naturally represented by adjacency matrix, but such matrices are generally sparse and high dimension, thus not well suited to statistical learning\cite{perozzi2014deepwalk}. How to represent network information in low dimension is an important task. There are classical methods of network representation which is dimension reduction based on calculating eigenvector, such as LLE\cite{roweis2000nonlinear, saul2000introduction}, Laplacian Eigenmap\cite{belkin2002laplacian,tang2011leveraging}, MDS\cite{cox2000multidimensional}, IsoMap\cite{tenenbaum2000global}, and DGE\cite{chen2007directed}. However, these methods do not perform well in large-scale networks.

\textbf{Representation Learning based on Deep Neural Network:}
In deep learning, more and more encoder-decoder models have been proposed to solve network representation problems. By optimizing a deterministic distance measure, those models can learn a node embedding from its neighbor nodes so as to project nodes into a latent space with a pre-defined dimensionality.

Recently, deep neural network\cite{collobert2008unified} based representation learning has been widely used in the natural language processing. Word2vec\cite{mikolov2010recurrent} is the deep learning model developed by Google to represent a word in a low dimension dense vector, which has  proven to be successful in natural language processing\cite{pennington2014glove}. By close analogy, topological paths neighboring a node may be handled like sequences of words, and word2vec can be adapted to network representation learning to reduce computing complexity and improve performance relative to conventional approaches. Accordingly, several recent publications have proposed word2vec-based network representation learning frameworks, such as DeepWalk\cite{perozzi2014deepwalk}, GraRep\cite{cao2015grarep}, TADW\cite{cao2015grarep}, CNRL\cite{tu2016community}, LINE\cite{tang2015line}, node2vec\cite{grover2016node2vec}, and metapath2vec\cite{dong2017metapath2vec}. All of the above frameworks utilize the skip-gram model\cite{mikolov2013efficient,mikolov2013distributed} to learn a representation of a node incorporating its topological context, so nodes with similar topological information will have similar numerical representations. Node representations are learned via skip-gram model by optimizing the likelihood objective using SGD with negative sampling\cite{levy2015improving}.

\textbf{Sampling Strategy:}
Similar to word sequences from documents, node sequences may be sampled from the underlying network as an ordered sequence of nodes\cite{dong2017metapath2vec}. Accordingly, different network representation learning frameworks adopt different node sampling strategies. DeepWalk\cite{perozzi2014deepwalk} deploys a truncated random walk to sample node sequences, and uses the skip-gram model to learn the representation of node sequences. However, DeepWalk only considers the first-order proximity between nodes. Moreover, it applies to unweighted networks. Practically, LINE is applicable for both weighted and unweighted networks and easily scales to large-scale networks with millions of nodes. The problem is that embedding of some loosely-connected nodes, which have few connected edges, heavily depends on their connected neighbors and unconnected negative samples\cite{xu2017empirical}. Most prior methods do not give full consideration to heterogeneity of  nodes and edges. Thus Deepwalk, LINE, and Node2vec are not effective for representing these heterogeneous networks. Sun et al.\cite{sun2011pathsim} introduced a metapath-based similarity measurement to find similar objects of the heterogeneous information networks. Furthermore, Dong et al. proposed metapath2vec\cite{dong2017metapath2vec} to capture heterogeneous structure and semantic correlation exhibited from large-scale networks by considering node types. However, one drawback of all previous methods is that they either only deal with homogeneous networks or do not consider edge semantics. When network contains nodes and edges with different types, the state-of-the-art embedding results are no longer effective as all of them do not consider edge semantics. To represent heterogeneity, we have developed edge2vec to learn node representations with general, systematic consideration of edge semantics.

\textbf{Representation learning in biomedical domains:}
In biomedical domains, there exist rich heterogeneous datasets about genes, proteins, genetic variations, chemical compounds, diseases, and drugs. Ongoing and expanding efforts to integrate and harness these datasets for data-driven discovery reflect widespread understandings of potential benefits to science and human health. For example,  Chem2Bio2RDF\cite{chen2010chem2bio2rdf} integrates over 25 different datasets related to drug discovery and comprises a large scale heterogeneous network. Such repositories hold complex relationships between many entity types. Representing this semantic complexity requires suitable embedding methods informed by these rich domains. Chen et al.\cite{chen2012assessing} propose Semantic Link Association Prediction (SLAP) to predict ‘missing links‘ between drugs and targets in Chem2Bio2RDF. Subsequently, Fu et al.\cite{fu2016predicting} applied an improved PathSim method with more than 50 metapaths on an extended Chem2Bio2RDF dataset to predict drug target interactions and rank metapaths based on Gini index. Although these previous studies focus on relationships between drugs and targets, none handle edge heterogeneity in graph embedding directly and generally. We regard this as an unmet need, both for the methodological value in exploring edge semantics, and also for the more practical value of applying and validating this novel methodology via biomedical data science use cases.  Thus motivated, we propose edge2vec as an improved representation learning model, well suited for discovery on biomedical knowledge graphs.

\textbf{Context: applied machine learning and data science:}
Machine learning is a big, diverse and rapidly advancing area of research, challenging to monitor and contextualize even to its scholars and practitioners. For biomedical data science, the challenge is compounded with applicability to complex real world datasets and tasks. This is applied machine learning, wherein the final evaluation depends on relevance and comprehensibility to the application domain. In this paper we strive to maintain this relevance and comprehensibility through concise, contextualizing notes and examples, with appropriate citations for further study. Terminology is a related challenge, for example the equivalence of "graph" and "network", so to assist we provide a glossary as supplementary material.

\section*{Methods}
\begin{figure}
\includegraphics[width=0.95\columnwidth]{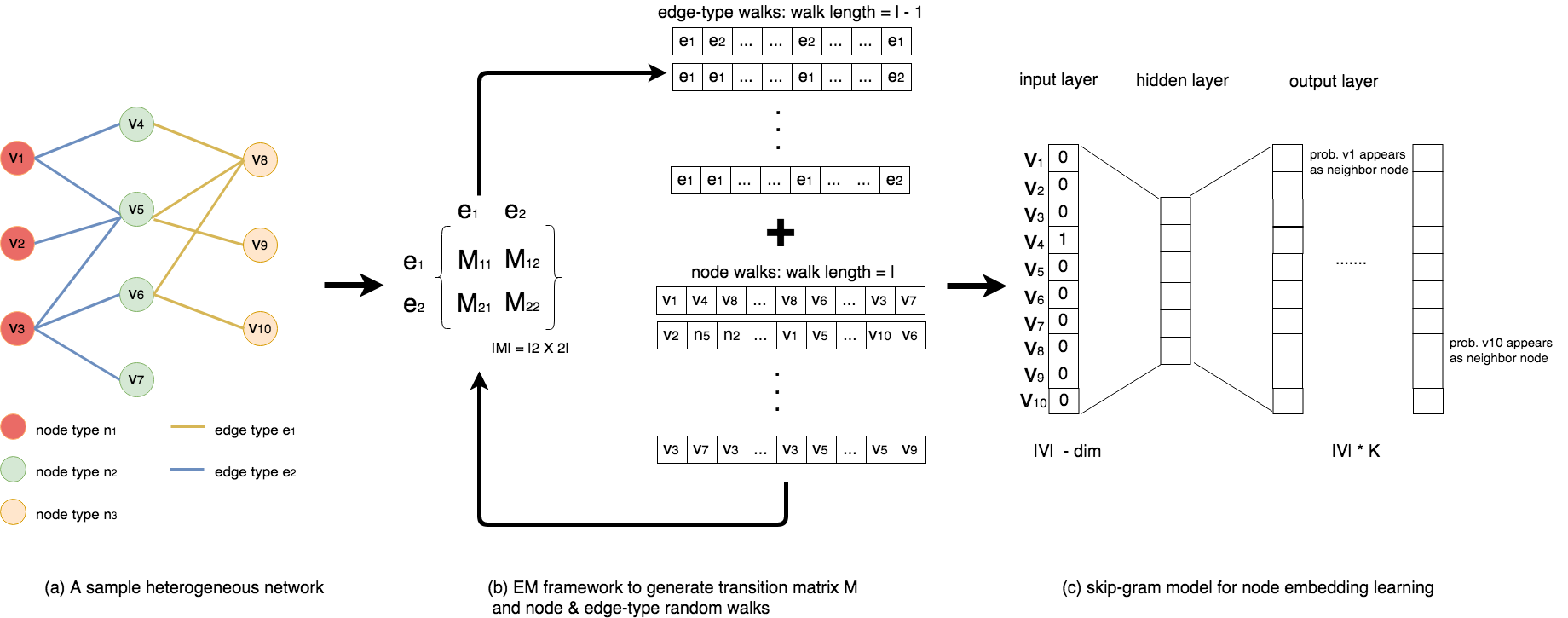}
\caption{ \csentence{An illustrative pipeline of edge2vec.}(a) a heterogeneous network with three types of nodes and two types of edges, colored by types. (b) EM framework to optimize an edge-type transition matrix $M$ and generate node random walks as well as related edge-type corpus. (c) skip-gram model is used for node embedding learning. For a node $v_{4}$, the input layer is its one-hot encoding and the output layer is the one-hot prediction for all its $K$ neighbor nodes (e.g. node $v_{1}$ and node $v_{10}$).} 
    \label{fig:pipline}
\end{figure}

In this section, we introduce edge2vec. The pipeline is shown in Figure \ref{fig:pipline}. We treat heterogeneous network embedding learning as an optimization problem and design an EM framework associated with a skip gram model to solve it. See Algorithm ~\ref{al:pseudo} pseudo code for details. 

\subsection*{Edge-type transition matrix for network embedding}

As word2vec\cite{mikolov2010recurrent} informed node2vec\cite{grover2016node2vec}, we can represent a node and its network neighborhood analogous to a word-context relationship in a text corpus. Random walk paths of nodes are akin to word sequences. We thereby convert the node embedding learning problem into a node neighborhood optimization problem: given a node, we need to maximize the probability of neighbor nodes, which is Formula 1:
\begin{equation}
  \argmax_{\theta} \prod_{v \in V}\prod_{c \in N(v)} p(c|v;\theta)
  \label{eq:hnode2vec}
\end{equation}


where V refers to the node collection of the network G(V,E); N(v) refers to the neighbor node collection of node v; $\theta$  is the node embedding parameterization to be learned. 

However, this optimization only works well in homogeneous networks. As in heterogeneous networks, different types of nodes and edges occur with varying frequency. But low frequency node and edge types may be very important, depending on their semantics in the knowledge domain. For instance, in a scholarly citation network, venue nodes (i.e., conferences and journals) are fewer but more important than publication nodes. Since node2vec would treat all nodes equally, knowledge contained in the venue relationships would be lost. Likewise, throughout biomedical domains, node and edge semantics must be considered to avoid loss of critical knowledge. For one example, the edge relationship between an approved drug and its well validated protein target is highly and exceptionally informative, reflecting prodigious research efforts and expense. To address this need for edge semantics, we design an edge-type transition matrix which holds the transition weights between different edge types during the random walk process. Therefore, we consider not only the topological structure of the network but also edge semantics. Accordingly, the optimized version is shown in Formula 2:
\begin{equation}
  \argmax_{\theta,M} \prod_{v \in V}\prod_{c \in N(v)} p(c|v;\theta;M)
  \label{eq:hnode2vec}
\end{equation}

M refers to the edge-type transition matrix. The matrix stores the random walk transition weights between different edge types. By employing the transition matrix as a prior distribution guiding the random walk process, we not only consider the distance between the next-step node and the previous-step node but also the weight between the next-step traversed edge type and the previous-step traversed edge type. Therefore, we can normalize by type so that the effect of low frequency node/edge types won't be lost by dilution among high frequency node/edge types. As shown above, the optimization function maximizes the probability of generating the node neighborhood of a given node v, thus the transition probability from the current node v to its neighbor c can be seen in Formula 3:


\begin{equation}
  p(c|v;\theta;M) = \frac{e^{\vec{f_{v}} \cdot \vec{f_{c}}}}{\sum_{u \in V}e^{\vec{f_{u}} \cdot \vec{f_{c}}}}
  \label{eq:prob}
\end{equation}
where $\vec{f_i}$ means the current step embedding for node \textit{i} which will be updated in each batch. We calculate the inner product of two node embeddings, which are normalized by a Softmax function.

We designed an EM framework to combine the update of the transition matrix M and optimization of node context into a unified framework. An edge-type transition matrix is initialized with all values set to 1, meaning initially, all edge type transitions are regarded as equally probable. Then, we iteratively generate the random walk corpus of paths, optimizing the transition matrix based on the sampled frequencies of edge type transitions.
\begin{algorithm}
\caption{edge2vec algorithm}
\begin{algorithmic} 
\REQUIRE Graph$<V,E>$ $g$, Edge-type transition matrix $M$ 
\STATE initialize $walks$ empty, all values in $M$ as $1$,  node embeddings $f$
\STATE $walks, M$ = GenerateTransitionMatrix($g,M$) 
\STATE $f$ = StochasticGradientDescent($walks$)

\STATE \textbf{return} $f$
\\\hrulefill

\STATE \textbf{GenerateTransitionMatrix}($g,M$)
\STATE initialize \# of iteration N
\WHILE{$N > 0$}
\STATE $N \leftarrow N - 1$
\STATE \#E step
\STATE $walks$ = HeteroRandomWalk($g,M$)
\STATE \#M step
\STATE $\vec{v_{i}}$ = vector with each dimension as the \# of edge type $i$ in each walk from $walks$
\STATE $M_{ij}$ = Sigmoid( PearsonCorrelation($\vec{v_{i}},\vec{v_{j}}$) )
\ENDWHILE
\STATE return $walks$, M
\STATE
\hrulefill
\STATE \textbf{HeteroRandomWalk}($g,M$)
\FOR{node $n \in g$} 
\STATE initialize an empty node walk $w$, an empty edge walk $T$, given random walk length $l$
\STATE Append $n$ to $w$
\WHILE{length(w) < $l$}
\IF{length$(w) == 1 $} 
\STATE $curr$ = $w[-1]$
\STATE Random sample node $m$ from Neighbour($curr$) based on edge weight. 
\STATE Append $m$ to $w$
\STATE Append EdgeType($curr,m$) to $T$ 
\ELSE
\STATE $curr$ = $w[-1]$, $prev$ = $w[-2]$
\STATE $p_{1}$ = T[-1]
\FOR{node $k \in$ Neighbour($curr$)} 
\STATE $p_{2}$ = EdgeType($k,curr$)
\STATE $EW(k,curr) = M[p_{1}][p_{2}] \cdot  W(k,curr) \cdot \alpha_{pq}(k,u)$ \#$W$(k,curr) is edge weight between node $k$ and $curr$.
\STATE Random sample node $m$ from Neighbour($curr$) based on updated edge weight $EW(k,curr)$. 
\STATE Append $m$ to $w$
\STATE Append EdgeType($curr,m$) to $T$ 
\ENDFOR
\ENDIF 
\ENDWHILE  
\ENDFOR
\STATE \textbf{return} $T$
 
\end{algorithmic}
\label{al:pseudo}
\end{algorithm}
\subsection*{Expectation-Maximization framework}
\subsubsection*{Expectation step}
Assume we have a set of E=$ \{ $ e\textsubscript{1}, e\textsubscript{2}, e\textsubscript{3}$ \ldots $  e\textsubscript{m}$ \} $  different edge types in a network. From the previous iteration in the EM framework, we can get a collection of random walk paths for each node as P = $ \{ $ p\textsubscript{1}, p\textsubscript{2}, $ \ldots $ p\textsubscript{n}$ \} $ . In each walk path p\textsubscript{i} (i $ \in $  $ \{ $ 1,2$ \ldots $ n$ \} $ ), it is constructed like p\textsubscript{i} = $ \{ $ n\textsubscript{1},n\textsubscript{2},n\textsubscript{3}, ...,n\textsubscript{l} $ \} $  where n\textsubscript{i} is the ith node in p\textsubscript{i} and l is a predefined walk length. Based on each path, we first extract all edges $ \{ $ T(n\textsubscript{1}, n\textsubscript{2}), T(n\textsubscript{2}, n\textsubscript{3}), $ \ldots $ , T(n\textsubscript{l-1}, n\textsubscript{l})$ \} $  in the path by locating every start node n\textsubscript{k} and end node n\textsubscript{k+1} where k $ \in $  $ \{ $ 1, 2, ..., l $-$  1$ \} $ , e\textsubscript{k} = T(n\textsubscript{i}, n\textsubscript{j}) refers to the edge type between n\textsubscript{i} and n\textsubscript{j}. After that, we calculate the number of times each type of edge e\textsubscript{j} (e\textsubscript{j} $ \in $  E) appears in the walk path p\textsubscript{i}. The same calculation is applied to all walk paths. In the end, for each edge type e\textsubscript{j}, we get a vector representation v\textsubscript{j}, where the ith dimension in the v\textsubscript{j} refers to the number of times e\textsubscript{j} appears in walk path p\textsubscript{i} . One assumption of our model is for a pair of edge type e\textsubscript{1} and e\textsubscript{2}, the distribution of each edge type sampled from the random walk paths is a valid estimator for the transition correlation for the graph. Hence, by calculating the correlation between their associated vector v\textsubscript{i} and v\textsubscript{j} in the walks, we can regard the correlation score as their updated transition weight. Therefore, we can define the formula for updating transition matrix as Formula 4:


\begin{equation}
  M(e_{i},e_{j}) = \text{Sigmoid}(\frac{E[\vec{(v_{i}}-\mu(\vec{v_{i}}))\vec{(v_{j}}-\mu(\vec{v_{j}}))]}{\sigma(\vec{v_{i}})\sigma(\vec{v_{j}})}) 
  \label{eq:update}
\end{equation}

where $E[\cdot]$ is the expectation value and $\sigma$ is related standard derivation value. M(e\textsubscript{i} , e\textsubscript{j}) refers to the updated transition weight between edge type i and j. v\textsubscript{i} and v\textsubscript{j} are vector representation of e\textsubscript{i} and e\textsubscript{j} on all walk paths. By using Pearson correlation analysis, we can get a pairwise correlation score between two edge types to check the distribution difference. Larger weight value means larger correlation between the pair of edge types. However, as the range of the correlation score varies from -1 to +1, it makes no sense if we keep the original negative weights between a pair of edge types. Because we involve the optimized transition weights to the random walk probability, and the probability can’t be negative, thus we normalize by transformation to a Sigmoid function to solve this issue and restrict the transition probability in a range of between 0 and 1. Moreover this non-linear transformation can better help to capture the patterns of transition probability than other linear or quadratic transformation functions\cite{kedem2012non}. The definition of Sigmoid$(\cdot)$ is shown as Formula 5:

\begin{equation}
  \text{Sigmoid}(x) = \frac{1}{1+e^{-x}}
  \label{eq:relu}
\end{equation}
In summary, the non-linear transformed correlation ensures three characteristics of the biased random walk on a heterogeneous network: First, a random walk tends to pass on edges with same edge-type. Based on the correlation calculation in Formula 4, given an edge-type, the correlation with itself is always +1, which is the highest transition weight inside the transition matrix. Second, the Sigmoid function guarantees optimization convergence of transition matrix. Transition weights are adjusted according to the Sigmoid function by training based on the correlations calculated from the random walks until a stable final value is reached. Third, only edge-types with closer relationships tend to have higher transition weights. Although some edge-types are \textit{globally} more common and likely to appear in random walks, we consider the \textit{specific} co-occurrence rates between edge-types in the same random walk. For example, if edge-type $a$ appears (20,30,40) times in three random walks, while edge-type $b$ appears (1,1,1) times in the same randoms walks. The transition weight from $b$ to $a$ is still considerably low.

\subsubsection*{Maximization step}
\begin{figure}
\includegraphics[width=1\columnwidth]{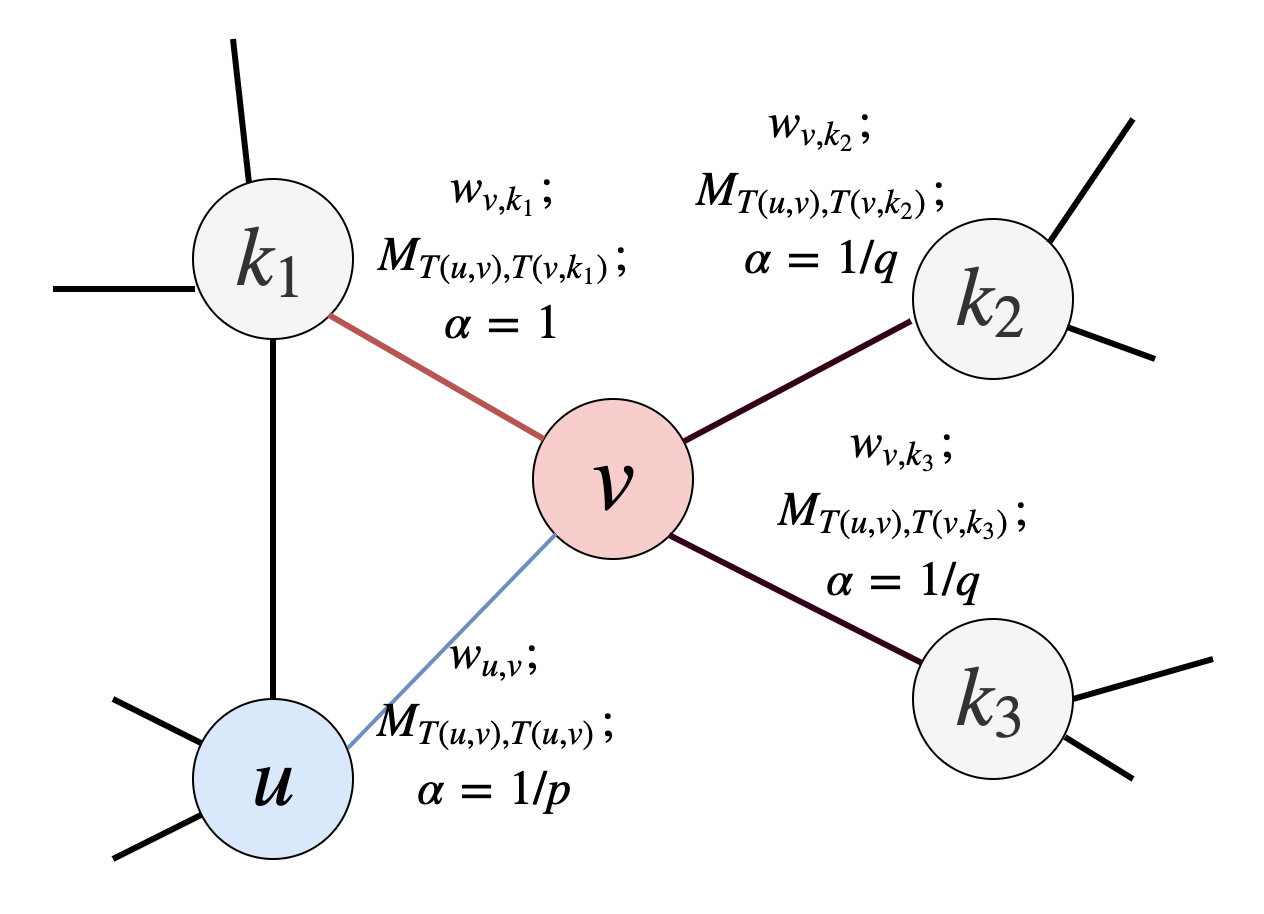}

\caption{Three parts of the weights to guide the biased random walk on heterogeneous networks.} 
    \label{fig:demo}
\end{figure}

In each iteration in the EM framework, based on the updated edge-type transition matrix M in the expectation step and the network topological structure, the biased random walk process generates a new paths with information of nodes and node neighbors. The transition matrix contributes to the calculation of random walk probabilities, thereby including the influence of edge-type information in sampling, which can reduce the negative effects caused by skewed type distribution issues. Even though some types of edges appear less frequently in the network, if the transition weights between those edge-types and other edge-types are high, the edge still has a high probability to get visited during the random walk process. Another important feature is that based on the expectation step, as well as Formula 4, for an edge-type e, $M_{e,e}$ is always the largest among all possible edge-type pairs toward e, which means random walk prefers to keep the same kind of edge-type. So, during the random walk process, given the current node v and the previous node u, the probability for the next candidate node n is calculated as Formula 6 and demonstrated in Figure \ref{fig:demo}:


\begin{equation}
  p(n|v;u;M) = \frac{w_{vn}\cdot M_{T(u,v)T(v,n)}\cdot \alpha_{pq}(n,u)}{\sum_{k \in N(v)}w_{vk}\cdot M_{T(u,v)T(v,k)}\cdot \alpha_{pq}(k,u)}
  \label{eq:random_walk_pr}
\end{equation}

where T(v,u) refers to the edge-type between node v and node u. $\alpha_{pq}(k,u)$ is defined based on the distance $d_{ku}$ between next step node candidate k and previous traversed node u. The distance function is defined as Formula 7:

\begin{equation}
  \alpha_{pq}(k,u) =
  \begin{cases}
    \frac{1}{p},       & \quad  d_{ku} = 0\\ 
    1,  & \quad d_{ku} = 1\\
    \frac{1}{q},  & \quad d_{ku} = 2\
  \end{cases}
  \label{eq:distance}
\end{equation}

As seen in Algorithm 1, at the beginning, we initialize walk paths as empty, all values in the transition matrix as 1, we use function $GenerateTransitionMatrix(\cdot)$ to utilize an EM framework to get walk paths and the matrix M. In maximization steps, the function takes transition matrix in the last iteration as input, invokes the $HeteroRandomWalk(\cdot)$ function to get walk paths, the probability of random walk is mainly based on Formula 6. In expectation steps, the function utilizes the updated walk paths to optimize the transition matrix by Formula 4. We can retrieve an optimized edge-type transition matrix, which holds the correlation between edge-types, via the EM framework. At the same time, we can also get the random walks as a node "corpus", which holds the correlation between nodes. We therefore represent the whole heterogeneous network as a collection of random walk paths, which can be used as the input of the next step for embedding optimization.

\subsection*{Skip gram for embedding optimization}
With the help of the EM framework, we can get the transition matrix M and random walks \textit{w} as the input layer to train the node embedding via a one layer neural network. To optimize the Formula 2, we use the stochastic gradient descent (SGD) method to get optimized node embeddings. Considering all nodes to maximize Formula 2 would be slow and computationally inefficient. Hence, in addition to the known neighbor node \textit{t}, we use the negative sampling method to generate \textit{k} negative nodes towards a given node \textit{v}. And the K negative nodes $u_{i}$ where $i \in \{1,2,...,k\}$ are randomly sampled from  the uniformed distribution $D(t)$ with probability $P(t)$. Moreover, we take logarithm on Formula 2 to reduce calculation complexity. And the final objective function turns to be Formula 8 in the end:

\begin{equation}
  \mathbf{O}(f) =  \text{log [Sigmoid}(\vec{f_{t}}^T \vec{f_{v}})] + \sum_{i=1}^{k}E_{u_{i} \sim P(t|t \sim D(t))}\text{log [Sigmoid}(\vec{-f_{u_{i}}}^T \vec{f_{v}})]
  \label{eq:model_objective_function}
\end{equation}

The goal of the objective function is to maximize the similarity with the positive neighbour node and minimize the similarity with negative neighbor nodes.
\section*{Results}
In this section, we describe the biomedical dataset used to test edge2vec and demonstrate the advantage of our model in three evaluation tasks. Moreover, we have a separate section for parameter tuning to retrieve the best model in both efficacy and efficiency points of view.

\subsection*{Biomedical dataset: Chem2Bio2RDF}
\begin{figure}
    \includegraphics[width=0.8\columnwidth]{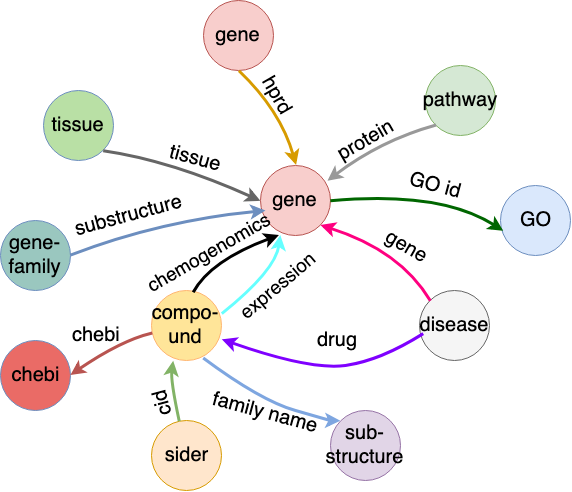}
    \caption{Chem2Bio2RF medical data graph structure.}
    \label{fig:graph_representation}
\end{figure}

Chem2Bio2RDF\cite{chen2010chem2bio2rdf} is a richly heterogeneous dataset integrating data from multiple public sources spanning biomedical sub-domains including bioinformatics, cheminformatics and chemical biology. The dataset includes 10 node types and 12 edge types. For details of each node/edge-type description, please refer to Table \ref{tab:data}. In total, there are 295,911 nodes and 727,997 edges, a relatively sparsely connected network. There exist multiple edge types between two given node types, for example, two edge types between node types "gene" and "compound." Node and edge type distributions are highly skewed. For instance, there are more than 20,000 compound nodes but a relative few are well studied in biological experiments, such as approved drugs, while most have few high confidence biological associations. Overall, the heterogeneity comprised by these network characteristics present significant challenges for embedding learning, and moreover, the particulars and specific semantics of this biomedical knowledge graph are essential considerations in optimizing learning power. Figure \ref{fig:graph_representation} shows the whole network structure of Chem2Bio2RDF.

 \begin{table}
        \centering
    \begin{tabular}{|l|l|l|l|l|l|l|} \hline
          
         \textbf{node type}  & \textbf{number}   & \textbf{edge type}  &  \textbf{number} & \textbf{edge type description}\\ \hline
           gene & 21,738      &  hprd  &  30,215 & protein protein interaction\\ \hline
           compound &  258,003    &  protein  & 11,258 & has pathway\\ \hline
            chebi&   2,777     & tissue   & 10,178 & tissue gene expression\\ \hline
           pathway &  192     &   GO id & 95,422 & has GO \\ \hline
           sider &  1,051  &      family name   & 7,181 & has gene family\\ \hline
            gene-family& 329   &    gene    & 2,929 & cause disease \\ \hline
           GO &  9,710  &  chebi    & 15,986 & has chemical ontology\\ \hline
           substructure & 290   &  drug    & 943 &has pathway \\ \hline
            tissue& 507   &  cid  &    9,004 & cause side effect\\ \hline
           disease &   1,284     & expression   &  16,167 & compound gene expression\\ \hline
            &    &  substructure  &  6,169  & has substructure  \\ \hline
            &    &chemogenomics    &  522,545 & bind  \\ \hline
         \end{tabular}
         \caption{Node and edge description in Chem2BioRDF }
         \label{tab:data}
 \end{table}

Given the proposed edg2vec, we set up parameters with p = q = 0.25; embedding dimension d = 128; for other parameters we use the defaults from node2vec. After those parameters are assigned, we use Chem2BioRDF to train our edge2vec model. To evaluate the fitness of the generated node embeddings, we propose three evaluation tasks in the following three sections.
\subsection*{Evaluation Metrics}
In this paper, we evaluate our model from both classification and information retrieval viewpoints. 

Precision, recall, F1 score and Hamming loss are four metrics reported in classification tasks. Precision implies  the ratio of correct positive results returned by the classifier; recall implies the ratio of correct positive results are returned; F1 score balances both precision and recall by taking their harmonic average. All above three metrics are in a range of 0 and 1, the higher the better. While the Hamming loss is the fraction of labels that are incorrectly predicted. The score is also in a range of 0 and 1, but the lower the better.

Precision@K, recall@K, MAP, NDCG and reciprocal rank are five metrics reported in information retrieval related tasks. Precision@K and recall@K imply the precision and recall score in the Top K ranked results. MAP refers to "mean average precision", which implies the average precision score for all searching queries. NDCG refers to "normalized discounted cumulative gain", which is a metric to measure not only the accuracy of searching results but also the ranked position of correct results. Like NDCG, reciprocal rank also considers the correct results ranking positions in the returned ranking list. It is the multiplicative inverse of the rank of the first correct result among all searching queries.

\subsection*{Entity multi-classification}
\begin{table}
        \centering
    \begin{tabular}{|l|l|l|l|l|l|} \hline
        \textbf{Algorithm} &  \textbf{Precision}       & \textbf{Recall}   &\textbf{F1 measure} & \textbf{Hamming loss}     \\ \hline
         DeepWalk  & 0.5624    & 0.5708  &  0.5650   & 0.4291    \\ \hline
         LINE  & 0.6366  &0.6390   & 0.6279  &   0.3609  \\ \hline
         node2vec  & 0.5652  & 0.5656  &  0.5622 &  0.4343   \\ \hline
         edge2vec        & \textbf{0.7554}* &\textbf{0.7546}*  &     \textbf{0.7544}*& \textbf{0.2453}*\\ \hline
         
         \end{tabular}
         \caption{Classification on node labels in the medical network. Symbol "*" highlights the cases where our model significantly beats the best baseline with $p$ value smaller than 0.01.}
         \label{tab:multiclass}
 \end{table}
We first propose a node multi-classification task. In this task, we take the types of nodes away so the network only has nodes, edges, and edge-types. We run edge2vec and cluster nodes based on the result of edge2vec to see whether nodes with similar types will be clustered together. In the Chem2BioRDF dataset, there are 10 different node types with different scale number. In order to build up a suitable dataset for the classification model, for each node type, we randomly sample equal number of nodes from the dataset. In this way, we have a natural baseline as precision = 0.1 for a random classifier. Each node is represented as an instance; the 128 dimension vectors are regarded as 128 different features. Its related node type is the response variable. We use a linear support vector machine as the classification model to predict the node’s labels, and use a 10-fold validation to evaluate the returned metrics. Three network embedding methods including DeepWalk, LINE and node2vec are our baseline algorithms. For node2vec, we take p = q = 0.25 which is the same setting as edge2vec. Other settings for all three algorithms are just default settings according to their related publications. For each node, after we learn its node embeddings for all baselines, we concatenate the embedding with the number of edges it has for each edge-type to integrate edge-type information into all baseline models as well. For example, if there are four edge-types in a network and a node has one edge with type 1, two edges with type 2, three edges with type 3 and zero edge with type 4, we concatenate an additional four dimensional vector (1,2,3,0) to the original learned embedding. As metapath2vec requires metapath definitions (manually curated) and thereby only uses selected metapath-pattern matched nodes for training node embeddings, metapath2vec is not comparable with other algorithms for a multi-classification task, which is also a drawback of metapath2vec. 

We use precision, recall, F1 score macro, and Hamming loss as four evaluation metrics. These are all commonly used evaluation metrics particularly for classification problem. Precision is the fraction of relevant instances among the retrieved instances, while recall is the fraction of relevant instances that have been retrieved over the total amount of relevant instances. F1 measure is the harmonic average of the precision and recall, which balances the two metrics. Hamming loss is the fraction of labels that are incorrectly predicted. Details of the evaluation results can be seen in Table \ref{tab:multiclass}. To verify our model's superiority, we run our model five times and calculate the performance differences between our model and the best baseline on each metric for all the runs, and apply a T-test to check whether the performance difference is significantly above 0 or not.

From the evaluation results, we can find all four algorithms can predict node types far better than a random classifier. It means even we treat this heterogeneous network as a homogeneous one, there is still some meaningful information stored in these node embeddings. DeepWalk and node2vec have similar results which is no wonder because DeepWalk can be regarded as a particular node2vec model when p = q = 1. While LINE performs the best among all three baselines. It means for this medical network, local structure (one step neighbours and two step neighbours contains most information of a node). However, our proposed edge2vec model outperforms all baseline algorithms,. In all four evaluation metrics, our model has at least 20$\%$  improvement in each evaluation metric. It reflects that our model can better predict node labels via its node embedding. Moreover, in all steps of edge2vec, we only use edge-type information during the random walk to generate edge-type transition metrics, and no node type information. Therefore, we can rigorously validate model performance in node type prediction.


\subsection*{Compound-gene bioactivity prediction}
 \begin{table}
        \centering

    \begin{tabular}{|l|l|l|l|l|l|l|} \hline
        \textbf{Algorithm} &  \textbf{Precision}       & \textbf{Recall}   &\textbf{F1 measure} & \textbf{Hamming loss}  & \textbf{AUROC}  \\ \hline
         DeepWalk  & 0.7787   & 0.7750 &  0.7742  &  0.2250  & 0.7660\\ \hline
         LINE  & 0.8170 & 0.8166 & 0.8166 &    0.1833 &  0.8058\\ \hline
         node2vec  &0.7983  & 0.7916 & 0.7904 &   0.2083   & 0.7793\\ \hline
         metapath2vec (Co-Ge-Co)      & 0.5170 & 0.5170 & 0.5168 & 0.4830  &  0.5007 \\ \hline
         metapath2vec (Co-Ge-Ge-Co)   & 0.4979 & 0.4980 & 0.4976 & 0.5020   & 0.4890   \\ \hline
         metapath2vec (Co-Dr-Ge-Dr-Co)    & 0.5305 & 0.5305       &  0.5304 &0.4695 & 0.5304 \\ \hline
          metapath2vec++ (Co-Ge-Co)     & 0.4969 & 0.4970 &  0.4965     &0.5030 & 0.4776\\ \hline
         metapath2vec++ (Co-Ge-Ge-Co)     & 0.4854 & 0.4855     & 0.4854  & 0.5145& 0.4776\\ \hline
         metapath2vec++ (Co-Dr-Ge-Dr-Co)          &0.5120 & 0.5120 &  0.5119 & 0.4880& 0.5102\\ \hline
         edge2vec          & \textbf{0.9017}* &  \textbf{0.9000}*   & \textbf{0.8998}* &\textbf{0.1000}*& \textbf{0.8914}*\\ \hline
         
         \end{tabular}
         \caption{Compound-gene bioactivity prediction. Symbol "*" highlights the cases where our model significantly beats the best baseline with $p$ value smaller than 0.01.}
        \label{tab:link_prediction}
 \end{table} 
  
\begin{figure}
    \includegraphics[width=0.8\columnwidth]{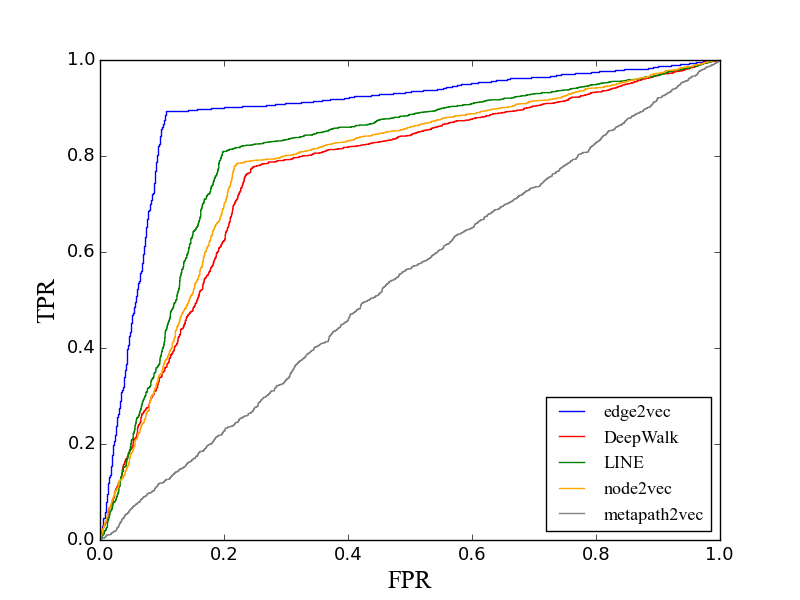}
    \caption{The ROC curve of compound-gene bioactivity prediction. the metapath2vec curve refers to the best result from all six reported metapath2vec/ metapath2vec++ models in Table \ref{tab:link_prediction}.}
    \label{fig:roc}
\end{figure}

One high-value biomedical challenge is to predict bioactivity between a compound and protein target (often referenced by the encoding gene). Such predictions can accelerate early stage drug discovery by informing and/or replacing expensive screening campaigns via virtual screening. Therefore, we considered the real world bioactivity prediction use case for a validation task. Besides the three baselines in the first task, we add metapath2vec to our baseline as well. As metapath2vec needs to define metapaths \textit{a priori}, three metapaths, which are compound-gene-compound; compound -gene-gene-compound; compound-drug-gene-drug-compound were selected as the metapaths for our baseline metapath2vec. As metapaths need to be symmetric, we have to take the network as undirected when training metapath2vec node embeddings. Our ground truth is from another work\cite{fu2016predicting} in which the authors generated 600,000 negative compound gene pairs and 145,6222 positive pairs. These ground truth pairs do not exist in Chem2BioRDF so it can be used as ground truth to evaluate of the result of edge2vec for bioactivity prediction. As the label for a compound-gene pair in ground truth is either ‘positive‘ or ‘negative‘, the prediction task is a binary classification task. Here, we randomly select 2,000 positive pairs and 2,000 negative pairs from the ground truth. And a random classifier will have an accuracy value as 0.5 naturally. Similar to the approach in the multi-class classification task, for each compound-gene pair, we use the difference of both embeddings together to form a new 128-dimension embedding to represent the pair, and we apply a logistic regression classifier to train a prediction model. Each dimension of the pair is also regarded as a feature. The relationship between the compound and gene is a binary label for each pair. In the training and testing process, If the prediction score is above 0.5, we label the pair as "positive", otherwise as "negative". We deploy the same evaluation metrics as  the multi-class classification task plus area under an ROC curve (AUROC) . The detailed result is shown in Table \ref{tab:link_prediction}. To verify our model's superiority,  we run our model five times and calculate the performance differences between our model and the best baseline on each metric for all the runs, and apply a T-test to check whether the performance difference is significantly above 0 or not. In Figure \ref{fig:roc}, we also report the ROC curve for edge2vec and baseline models based on their prediction scores, where we can find our model curve significantly performs better than the baselines.

Some interesting findings are observed from the experiments. First of all, among all three well known baseline algorithms (DeepWalk, LINE and node2vec), LINE still outperforms the other two baselines. And the result of DeepWalk is similar to that of node2vec. So, edge2vec is reliable and functionally stable for the two tasks. For metapath2vec, we leverage both metapath2vec and metapath2vec++ in our baseline models. As metapath2vec relies too much on selected metapaths, none of the three metapaths performs well.

Among these three metapaths, we find metapath compound-drug-gene-drug-compound works the best, implying that prediction accuracy is improved by the additional node types. Although the number of drug nodes is trivial compared with the number of compound nodes and gene nodes, drug nodes have larger effects than compounds and genes in terms of bioactivity prediction. So it is necessary to treat different types of nodes separately within an embedding model. Compared with metapath2vec, metapath2vec++ however achieves worse result in all three metapaths. edge2vec outperforms all baseline models. The F1 measure is around 0.9 which is far better than the random classifier with a score of 0.5. Also it has around 10$\%$  improvement compared to the LINE result which is the best of baseline results.

\subsection*{Compound-gene search ranking}
Bioactivity prediction as a binary classification task, like single point high throughput screening in the wet lab, predicts active or inactive only. This is helpful, but more useful is the capability to predict ranking of hits by a measure that increases the probability of success and overall efficiency in costly follow up efforts. Hence, this comprises our final evaluation task: compound-gene search ranking. By analogy, the number of hits returned by a search algorithm is generally less important than the ranking and particularly the top ranked hits. Thus, our final task can be described as an information retrieval or search efficiency task. To limit runtime cost, from the ground truth, we select 70 compounds, which contain more than one positive pair with a gene. For each compound, we calculate the top 100 similar nodes in Chem2BioRDF. Compared with the positive pairs of 70 compounds in ground truth, we evaluate the searching result using metrics such as precision, recall, MAP, NDCG, and mean reciprocal rank (MRR). These metrics care not only whether the bioactive genes are returned but also the ranking of the returned genes. For a compound node, if its bioactive genes shows up with a higher rank in the top 100 returned ranking list, the evaluation metrics will have larger values. After the bioactivity prediction task, we choose the best metapath among all three metapaths, which is compound-drug-gene-drug-compound. We evaluate the embedding results from LINE, node2vec, and edge2vec, as well as metapath2vec with the best metapath. Evaluation details are shown in Table \ref{tab:search}. To verify our model's superiority,  we run our model five times and calculate the performance differences between our model and the best baseline on each metric for all the runs, and apply a T-test to check whether the performance difference is significantly above 0 or not. From the evaluation table, we find DeepWalk and node2vec still have similar results, and both outperform LINE. metapath2vec is almost meaningless because all returned evaluation metrics are approaching to 0, which means it can barely retrieve future positive genes to compounds. Metapaht2vec++ performs slightly better than its previous performance in other two tasks and is comparable to LINE. And overall, node2vec works the best in all baseline algorithms. Compared with all baseline algorithms, our edge2vec outperforms all the baseline methods. Although the retrieved scores are all relatively small, there is around 10$\%$  improvement in precision and a little better in the rest of evaluation metrics at least. This, edge2vec adds value in this critical task of compound-gene ranking, which can improve cost efficiency in virtual screening follow up efforts of early stage drug discovery.

 \begin{table}
        \centering
    \begin{tabular}{|p{2cm}|l|l|l|l|l|l|l|l|} \hline
        \textbf{algorithm} &  \textbf{P@10} &  \textbf{P@100} &\textbf{Recall@10}  & \textbf{Recall@100}  & \textbf{MAP}   &  \textbf{NDCG}  &  \textbf{MRR}      \\ \hline
       
          DeepWalk    &0.0623  & 0.0198 & 0.0725 &  0.2780   &   0.0707&  0.1444   & 0.1502  \\ \hline
          
          LINE  & 0.0186 &    0.0069  & 0.0109  & 0.0360 & 0.0042  & 0.0234    & 0.0532 \\ \hline 
          
          node2vec   &   0.0714  & 0.0277 &   0.0859 & 0.2804 & 0.0786  &  0.1690   &   0.1676  \\ \hline
          
          metapath2vec (Co-Dr-Ge-Dr-Co)  &  0.0000 & 0.0001  & 0.0000  & 0.0011   & 0.0000  &   0.0004  & 0.0001 \\ \hline 
          
         metapath2vec++ (Co-Dr-Ge-Dr-Co) & 0.0157  & 0.0039 & 0.0082  & 0.0200  & 0.0040  & 0.0172 &  0.0509\\ \hline
         
          edge2vec  &  \textbf{0.0843}* &\textbf{  0.0329}*  &  \textbf{0.0990}* & \textbf{ 0.3092}*  &  \textbf{0.0809} &   \textbf{0.1840}*  & \textbf{0.1882}* \\ \hline
         \end{tabular}
         \caption{Searching  accuracy  for  retrieving  potential compound-gene bindings. Symbol "*" highlights the cases where our model significantly beats the best baseline with $p$ value smaller than 0.01.}
         \label{tab:search}
         
 \end{table} 
\subsection*{Parameter tuning}
In our EM framework for edge transition matrix training, in order to get the best fit to retrieve transition relationships between edge types, we have tried various correlation methods including cosine, Spearman, Pearson and Wilcoxon signed-rank. To standardize the returned correlation scores into a reasonable (above 0) and comparable range, we have tried various activation function such as ReLU, Sigmoid and traditional standardization methods. Our experiments show that using the combination Sigmoid activation function and Pearson correlation similarity performs the best to represent the transition relationship between edges. Hence, we decided to use this combination for building up our framework. During the transition matrix training process, there are four important parameters to be tuned. We list them with default value below.

\begin{enumerate}
\item{Number of walks on per node, r = 1}
\item{Walk length in each random walk path, w = 50}
\item{The ratio of nodes sampled for training edge transition matrix, p = 0.01}
\item{The number of iterations for training edge transition matrix, N = 10}
\end{enumerate}

\begin{figure}
\includegraphics[width=1.0\columnwidth]{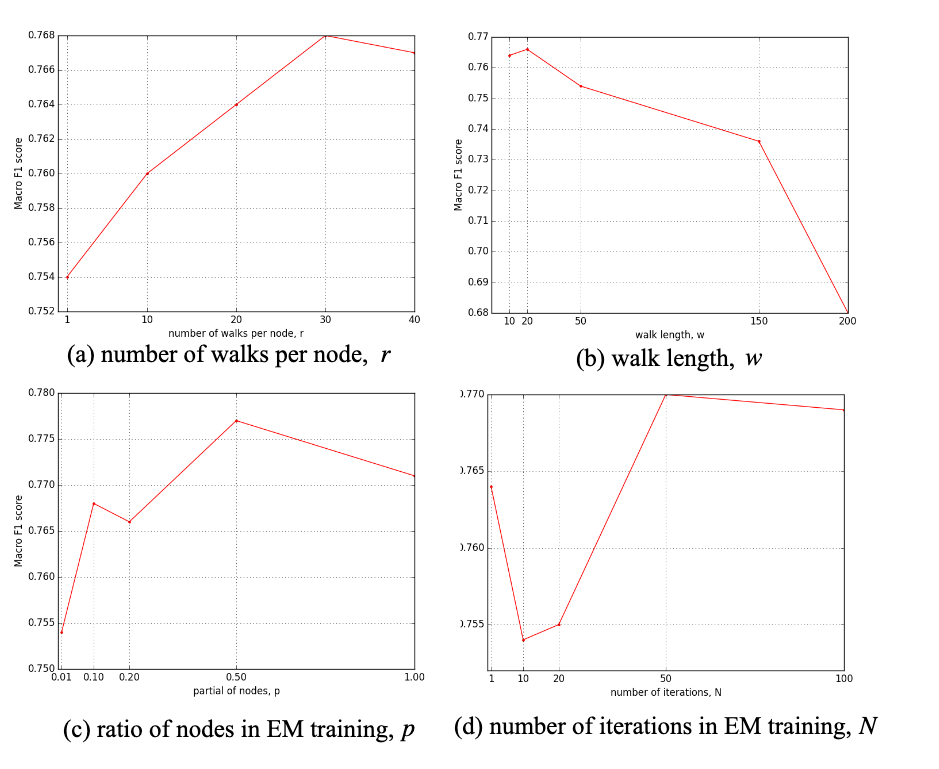}
    \caption{Parameter tuning in multi-class node classification.}
    \label{fig:tune}
\end{figure}

The default parameter settings are used to train our edge2vec model and compare with baseline models in previous sections. In this section, we vary each of them and fix the rest to examine the parameter sensitivity of our model. We leverage all generated results on solving node multi-class classification task and use the Macro F1 score as the judgment to evaluate related models. The result of our tuning process is shown in Figure \ref{fig:tune}.

To test how much that numbers of walk per node can affect our model, we test five cases r = $\{1,10,20,30,40\}$  and the result is in Figure \ref{fig:tune} (a). We can see that more numbers of walks on per node leads to an increase in Macro F1 score. The reason might be that more walks on a node can better reflect the edge relationships around the node and avoid the negative influence of walk randomness. However, even though it shows a positive trend, the increase is small. So a short number of walks on per node should be able to capture enough edge relation information around the node.

In Figure \ref{fig:tune} (b), with the increase of walk length, the Macro F1 score increases in the beginning and decreases later on. In general, as the random walk length increasing, it will tend to contain all types of edges. As we don't consider the sequence of the edge types in the same walk, increasing walk length can add noise obfuscating edge type transition relationships.

Figure \ref{fig:tune} (c) shows the different ratio of nodes involved in the EM training process to generate edge transition matrix. It might be no need to involve all nodes when training the edge transition matrix as long as all edge types can be covered in random walks in each iteration. Although with more nodes involved, the overall trend of Macro F1 score has a positive sign, the increase of Macro F1 score is not huge and the trend even fluctuates a little bit. We thereby conclude that using a sampling of nodes to train the edge type transition matrix is sufficient and computationally efficient.

Figure \ref{fig:tune} (d) refers to the influence of number of iterations in the EM process for edge type transition matrix training. We believe the reason why when N = 1 the Macro F1 score outperforms than  N = 10 is by chance as when we increase the number of iterations, the overall trend of Macro F1 score also increases. From the Figure \ref{fig:tune} (d), the convergence is fast, and a few iterations can already generate a satisfactory Macro F1 score.

\section*{Discussion}

To discuss how we can apply our node2vec model on a biomedical data set, we conduct two case studies to show the practical values of our model. The first one is to rank the positive links between nodes, which can be used for similarity search and recommendation; the second one is to cluster and visualize similar gene nodes that belong to the same gene family.

\subsection*{Ranking positive bindings for similarity search}
 \begin{table}
        \centering
    \begin{tabular}{|l|l|p{0.9cm}|p{0.9cm}|l|l|p{0.9cm}|p{0.9cm}|} \hline  
          drug & gene & node2vec similarity    & edge2vec similarity & drug & gene & node2vec similarity    & edge2vec similarity \\ \hline
         DB11348  &  CADPS2  &  0.1196  &  0.6199 &  DB09130 &    CLEC3B&  0.2177  &  0.6299  \\ \hline
         DB11348  & NUCB1   & 0.2221   & 0.6613 &  DB09130  &  HSPA13  & 0.1547   & 0.6278      \\ \hline
         DB11348 &  TPT1  &  0.2233  & 0.6439  &   DB09130  & IGLL1   &  0.1806  &  0.5856    \\ \hline
         DB11348  &  NUCB2  &0.2668    & 0.6867  &  DB09130  &  SFPQ  &  0.1275  &  0.5281      \\ \hline
         DB11348 &  CIB2  &  0.2761  &  0.6946   &   DB09130 &  A1BG  & 0.2478   &   0.6483    \\ \hline
         DB11348 &  CAPS  &  0.2827  &    0.6765&   DB08818 & HMMR & 0.1624  & 0.6072       \\ \hline
         DB11348 &  PEF1  &  0.3357  &  0.7472 &    DB08818  & CD44   &  0.1408  & 0.5670      \\ \hline
         DB11348 & CALR3   &  0.2396  &  0.6429 &   DB08818   &   HAPLN1 &  0.1672  &   0.5790     \\ \hline
         DB11348 & CADPS &  0.1715 &  0.5657  & DB08818 & VCAN   & 0.1877   & 0.6186         \\ \hline
         DB11348  &  FBN3  &  0.3306  &  0.7369 & DB09131  &  CSTB  &0.1675    & 0.5999    \\ \hline
         \end{tabular}
         \caption{Compare node2vec \& edge2vec difference on classification tasks}
         \label{tab:intro_case}
 \end{table} 
To verify how well our model can be used for similarity search and recommendation use cases, we carried out a ranking experiments using the links identified in the existing network. We randomly selected three widely used drugs from the network, which are Hyaluronic acid (DB08818), Calcium Phosphate (DB11348), Copper (DB09130), and Cupric Chloride (DB09131). Each of them has multiple target genes to interact with. The selected pairs of drugs and target genes exist in the network, and we want to reproduce the links using the cosine similarity score based on the embedding vectors. As we can see, our proposed edge2vec embedding can represent node similarity significantly better than the node2vec embedding. The cosine similarity scores for the drug targets of calcium phosphate were all above 0.6, indicating strong similarity between the drug and the target genes. However, using node2vec embedding the cosine similarity scores between calcium phosphate and its targets were all below 0.4, and some of them demonstrated strong dissimilarity like CADPS2 as a target gene of calcium phosphate. The same findings for the other three drugs and their target genes. In addition, all of the target genes for those drugs can be identified as similar nodes with high rankings using edge2vec embeddings. Details are shown in Table \ref{tab:intro_case}. we further performed a pairwise t-test\cite{box1987guinness} study to see whether the similarity scores generated by two models are significantly different or not. If edge2vec has significantly higher similarity score than node2vec, it means our model can better predict those positive drug-target pairs in the network. In our result, the difference between two embedding approaches is 0.0103 with a p-value of 0.0001. It means our embedding approach can better rank and retrieve the existing links in the network than node2vec.

\subsection*{Gene clustering analysis}
\begin{figure}
    \includegraphics[width=1\columnwidth]{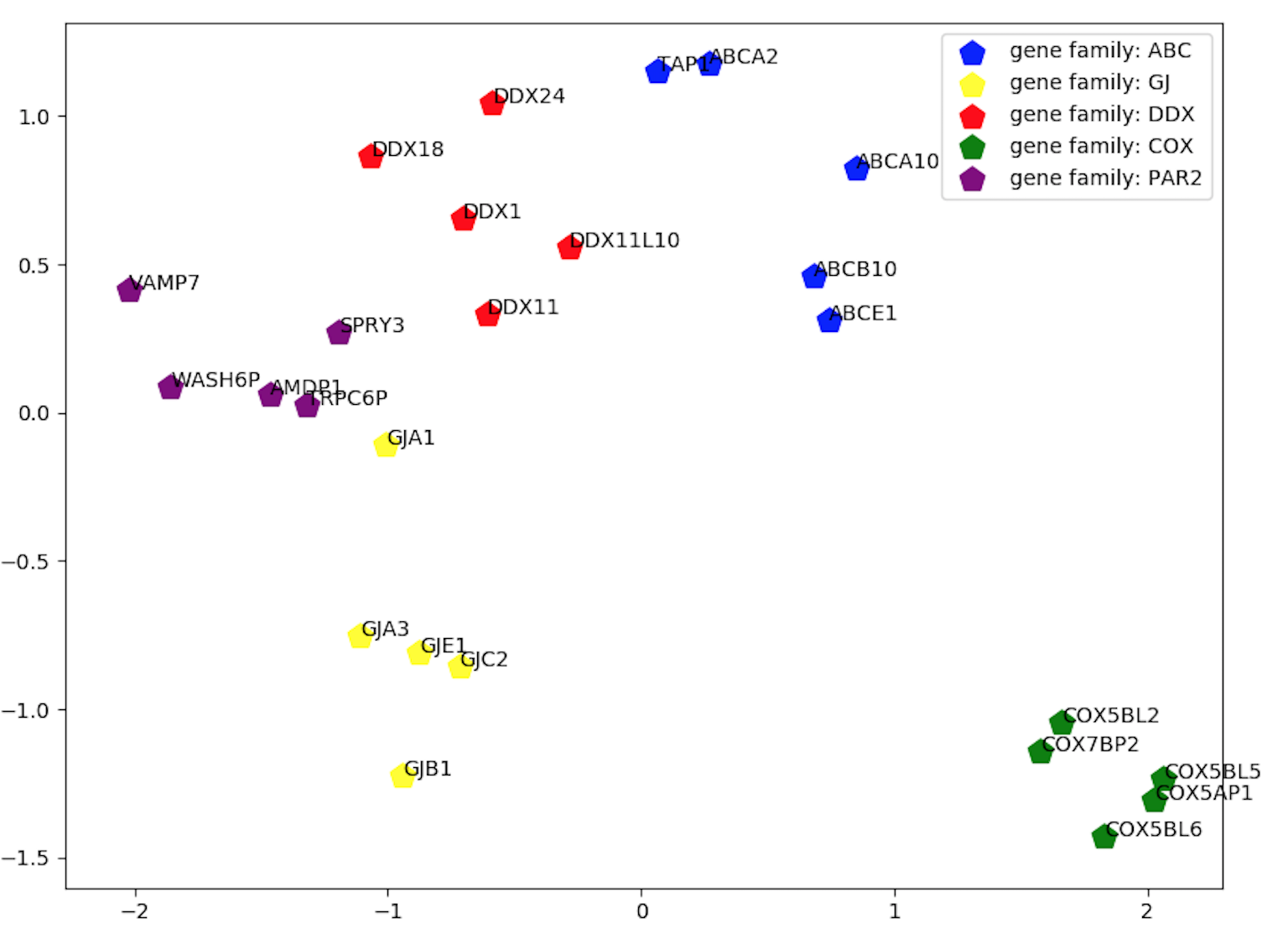}
    \caption{2-D PCA projection on 25 random selected genes, five each from 5 random gene families.}
    \label{fig:case_visualization}
\end{figure}
In order to further prove the usefulness of our node embedding results, we carried out a clustering analysis of gene nodes. We arbitrarily selected five gene families, which are ATP binding cassette transporters (ABC), Gap junction proteins (GJ), DEAD-box helicases (DDX), Cytochrome C Oxidase Subunits (COX), and Pseudoautosomal region 2 (PAR2). Each gene family refers to a collection of genes originated from the same root gene and performing similar biochemical functions. From each given gene family, five genes are randomly selected to perform clustering analysis. Then, we use principal component analysis (PCA) to project the default 128 dimensions of all gene embeddings into a 2-D space. Ideally, genes belonging to the same family should locate closer to each other than genes from different families. The resulting 2-D plot has shown that the twenty five genes in five gene families have been well clustered based on their node embedding vectors. Results can be visualized in Figure  \ref{fig:case_visualization}, where each pentagon refers to a gene and different colors indicate different gene families. It is easy to observe that genes are clustered by family in the 2-D space. In the 2-D projection, genes in family COX are all located in the bottom right corner of the plot, and genes in family GJ tend to stay in the bottom left corner. In a word, our edge2vec model can generate node embeddings highly reflecting their family information.

\section*{Conclusions}
In this paper, we propose edge2vec, which incorporates edge semantics to add value over previous methods, as evaluated by knowledge discovery tasks in the domain of biomedical informatics. Edge2vec employs an EM framework associated with a one-layer neural network, to learn node embeddings which perform better than previous methods for heterogeneous networks. The novelty of our work is to generate an edge-type transition matrix so that during the process to generate the node random walk corpus, heterogeneity of the network is also considered. It can reduce the skewed type distribution issue via weighted sampling. Moreover, compared with other state-of-art heterogeneous network embedding methods such as metapath2vec, our edge2vec has no restrictions and can deal with the situation where there are multiple relationships between two types of nodes. To illustrate efficiency and accuracy of our proposed model, we evaluate it on biomedical dataset Chem2BioRDF and propose three evaluation tasks including node multi-class classification, link prediction, and search rank efficiency. Edge2vec outperforms all baseline algorithms significantly. Furthermore, we illustrate the effect of edge2vec in biomedical domains using two case studies to explain the biological meanings of the prediction. Moreover, edge2vec can work well in both undirected and unweighted networks, and computational cost is only moderately increased relative to node2vec by choice of appropriate random walk strategy in the EM framework.

There are certainly promising future directions to be explored, which can be subdivided into (1) algorithmic modifications applicable to heterogeneous graphs generally,and (2) domain knowledge based enhancements applicable to characteristics of biomedical datasets and use cases. Informed by metapath2vec, we could change the objective function by using a node-type based negative sampling. Instead of random sampling from all types of nodes, we could sample negative nodes based on the ratio of each node type. Another opportunity for improvement involves adding domain knowledge into our existing model. During the random walk generation process, we have already considered both node distance (p,q) and edge-type (transition matrix M). In addition to these two attributes, we could add some pre-defined rules guiding random walks based on our domain knowledge. Another problem for such complex networks is the scale free issue, i.e. skewed degree distributions, where a relative few hub nodes account for the bulk of connections. To reduce this hub node effect in the network, we can also come up with new methods such as novel restriction rules in the random walk, or mitigate those effects by adding node degree related information to optimize a new objective function. For the node attribute prediction task, we can actually apply a semi-supervised approach: Given a sampling of nodes with known attributes, we can add this information into our embedding learning process and predict the attributes of remaining nodes. Or we can apply an unsupervised  approach: given the relationship between the target attribute with other known attributes, we use this relationship information and other known attributes information in our embedding learning process, and in the end directly predict node target attributes. These enhancements could extend our edge2vec model to better address specific downstream tasks such as node attribute prediction with unsupervised or semi-supervised training strategies.

\section*{List of Abbreviations}

\begin{table}
\centering
\begin{tabular}{|p{2.5cm}|p{9cm}|}
\hline
\textbf{Knowledge graph} & 
Knowledge base represented as nodes as entities and edges as relationships. A.k.a. Knowledge network. A type of heterogeneous graph. \\
\textbf{Machine learning} & 
Computer system automatically learns the data pattern via designed algorithms and statistical models.\\
\hline
\textbf{Heterogeneous graph} & 
Includes nodes of multiple classes. A bipartite graph has two classes, e.g. persons and movies. \\
\hline
\textbf{Graph embedding} & Transformation into feature space represented as numerical vectors. Node embedding is a form of graph embedding, where each vector represents one node. \\
\hline
\textbf{Adjacency matrix} & Graph topology square matrix, one row/column per node, typically very sparse.  \\
\hline
\textbf{Semantics} & 
Ontology defining types of entities and their relationships, in graph terms, nodes and edges, or in RDF terms, classes of entities and predicates. Formalized or not, the ontology conveys the semantics mapping data to knowledge. \\
\hline
\textbf{EM}& 
Expectation Maximization \\
\hline
\textbf{SGD} & 
Stochastic Gradient Descent \\
\hline
\textbf{word2vec} & 
ML feature method developed by Google representing a word and its lexical context in a low dimensional dense vector.\\
\hline
\textbf{node2vec} & 
Conceptually derived from word2vec, represents a node and its topological context. A path is treated like a sentence. \\
\hline
\textbf{Skip gram model} & Employed by word2vec, Node2vec and related methods, to represent non-adjacent words/nodes within some context window. \\
\hline
\textbf{Metapath} & Path pattern, the classes of nodes and edges, for which many instances may exist. \\
\hline
\textbf{metapath2vec} & 
Like node2vec but with heterogeneous graphs. Metapath based random walks. \\
\hline
\textbf{Transition matrix} & In the edge2vec algorithm, matrix of relative probabilities or expectation values for edge type transitions. \\
\hline
\textbf{Representation learning} & 
Processes by which an input representation is transformed to an alternative representation, such as in layers of a neural network. \\
\hline
\textbf{Biomedical Knowledge Domain} & 
Shorthand for a varied set of richly heterogeneous subdomains, including medical and pharmaceutical informatics, bioinformatics, and cheminformatics. \\
\hline
\end{tabular}
 \end{table}

\section*{Declarations}
\subsection*{Acknowledgements}
Not applicable.
\subsection*{Funding}
The work was supported by National Natural Science Foundation of China (No. 71573162). The funding bodies had no role in the design of this study,
the collection, analysis, and interpretation of data, or the writing of this manuscript.
\subsection*{Availability of data and materials}
The dataset used in this work can be found at Github \footnote{https://github.com/RoyZhengGao/edge2vec}. There are three files within the zipped folder: \textit{chem2bio2rdf.txt} is the heterogeneous graph we used to train our model. In this file, each line is a RDF triplet which contains two entities and their relations. Entity type and relation type can be obtained from their RDF representations directly. \textit{negative.txt} and \textit{positive.txt} stored the negative and positive bindings between genes and compounds, which are all directly generated from the original chem2bio2rdf paper. 

The dataset is processed from the original dataset published at BMC Bioinformatics \footnote{https://bmcbioinformatics.biomedcentral.com/articles/10.1186/s12859-016-1005-x}. We use the \textit{chem2bio2rdf.txt}, \textit{internal\_testset\_label/positive.txt} and \textit{internal\_testset\_label/negative.txt} from  \textit{semantic\_network\_dataset.zip} directly.

The source code is stored at at Github \footnote{https://github.com/RoyZhengGao/edge2vec}. For details to run the code, please refer to the instructions in the Github link. In order to load the data into edge2vec, please convert all RDF format data points to hashed ids first, and save in csv format. Then run \textit{transition.py} to generate and store the transition matrix. Then, \textit{transition.py} is utilized to load the data and transition matrix for embedding optimization.

For evaluation, please use \textit{negative.txt} and \textit{positive.txt} as ground truth files.
\subsection*{Authors' Contributions}
GZ prepared the dataset, programmed the algorithm, and initially drafted the manuscript. XL and YD conceived and guided the project. JJY revised the manuscript with emphasis on biomedical data science applications. GF helped to prepare the ground truth data set and comparison and helped with case study and editing. ST helped to implement the baselines. CO made contributions on the background writing. CG, BF, DW and QY are discussed the scientific ideas, reviewed and approved the final manuscript.

\subsection*{Ethics approval and consent to participate}
Not applicable.

\subsection*{Consent for publication}
Not applicable.

\subsection*{Competing Interests}
The authors declare that they have no competing interests.



\begin{backmatter}




\bibliographystyle{bmc-mathphys} 
\bibliography{Edge2vec_manuscript}      








\end{backmatter}
\end{document}